\documentstyle[aps,prl,epsf]{revtex}
\begin{document}

\title{Typical Performance of Gallager-type Error-Correcting Codes}
\author{Yoshiyuki~Kabashima$^{1}$, Tatsuto Murayama$^{1}$ and David~Saad$^{2}$}
\address{$^{1}$ Department of Computational Intelligence and Systems Science,
Tokyo Institute of Technology, Yokohama 2268502, Japan.  \\
$^{2}$The Neural Computing Research Group, Aston
University, Birmingham B4 7ET, UK.}
\maketitle
\begin{abstract}

The performance of Gallager's error-correcting code is investigated
via methods of statistical physics. In this
approach, the transmitted
codeword comprises products of the original message bits selected by
two randomly-constructed sparse matrices; the number of non-zero
row/column elements in these matrices constitutes a family of
codes. We show that Shannon's channel capacity is saturated for many
of the codes while slightly lower performance is obtained for others
which may be of higher practical relevance. Decoding aspects are
considered by employing the TAP approach which is identical to the
commonly used belief-propagation-based decoding.

\end{abstract}

\newcommand{\btau}{\mbox{\boldmath{$\tau$}}}
\newcommand{\bxi}{\mbox{\boldmath{$\xi$}}}
\newcommand{\bS}{\mbox{\boldmath{$S$}}}
\newcommand{\bz}{\mbox{\boldmath{$\zeta$}}}
\newcommand{\bJ}{\mbox{\boldmath{$J$}}}
\newcommand{\JO}{\mbox{\boldmath{$J^{0}$}}}
\newcommand{\cD}{{\cal D}}
\newcommand{\cJ}{{\cal J}}
\newcommand{\dashed}{\mbox{-\; -\; -\; -}}
\newcommand{\dotted}{\mbox{${\mathinner{\cdotp\cdotp\cdotp\cdotp\cdotp\cdotp}}$}}
\newcommand{\full}{\mbox{------}}

The ever increasing information transmission in the modern world is
based on communicating messages reliably through noisy transmission
channels; these can be telephone lines, deep space, magnetic storing
media etc.  Error-correcting codes play an important role in
correcting errors incurred during transmission; this is carried out by
encoding the message prior to transmission,
and decoding the corrupted
received codeword for retrieving the original message. In
his ground breaking papers, Shannon\cite{Shannon} analyzed the
capacity of communication channels, setting an upper bound to the
achievable noise-correction capability of codes, given their code (or
symbol) rate. The latter represents the ratio between the number of
bits in the original message and the transmitted codeword.

Shannon's bound is non-constructive and does not provide explicit
rules for devising optimal codes. The quest for more efficient codes,
in the hope of saturating the bound set by Shannon, has been going on
ever since, providing many useful but sub-optimal codes.

One family of codes, presented originally by Gallager\cite{Gallager},
attracted significant interest recently as it has been shown to
outperform most currently used techniques\cite{MacKay}. In fact,
irregular versions of Gallager-type codes have recently been shown to
get very close to saturating Shannon's bound in the case of
infinitely long messages\cite{Richardson}.  Gallager-type codes are
characterized by several parameters, the choice of which defines a
particular member of this family of codes. Most studies of
Gallager-type codes conducted so far have been carried out via
numerical simulations. Some analytical results have been obtained via
methods of information theory~\cite{MacKay}, setting bounds on the
performance of certain code types, and by combinatorical/statistical
methods~\cite{Richardson}; no quantitative results have been
obtained for their {\em typical} performance.

In this Letter we analyze the typical performance of Gallager-type
codes for several parameter choices via methods of statistical
mechanics. We then validate the analytical solution by comparing the
results to those obtained by the TAP approach to diluted systems and
via numerical methods.

In a general scenario, a message represented by an $N$ dimensional
Boolean/binary vector $\bxi$ is encoded to the $M$ dimensional vector
$\JO$ which is then transmitted through a noisy channel with some
flipping probability $p$ per bit (other noise types may also be
considered but will not be examined here). The received
message $\bJ$ is then decoded to retrieve the original message.

One can identify several slightly different versions of Gallager-type
codes. The one used in this Letter, termed the MN code\cite{MacKay} is
based on choosing two randomly-selected sparse matrices $A$ and $B$ of
dimensionality $M\!\times \!N$ and $M\!\times\! M$ respectively; these
are characterized by $K$ and $L$ non-zero unit elements per row and
$C$ and $L$ per column respectively. The finite, usually small,
numbers $K$, $C$ and $L$ define a particular code; both matrices are
known to both sender and receiver. Encoding is carried out by
constructing the modulo 2 inverse of $B$ and the matrix $B^{-1}A$
(modulo 2); the vector $\JO \! =\! B^{-1}A \ \bxi$ (modulo 2, $\bxi$
in a Boolean representation) constitutes the codeword. Decoding is
carried out by taking the product of the matrix $B$ and the received
message $\bJ\! = \!  \JO \! +\! \bz$ (modulo 2), corrupted by the
Boolean noise vector $\bz$, resulting in $A\bxi\! + \! B\bz$. The
equation
\begin{equation}
\label{eq:decoding}
A\bxi + B\bz = A\bS + B\btau
\end{equation}
is solved via the iterative methods of Belief Propagation
(BP)\cite{MacKay} to obtain the most probable Boolean vectors $\bS$ and
$\btau$; BP methods in the context of error-correcting codes have
recently been shown to be identical to a TAP\cite{tap} based solution
of a similar physical system\cite{us_sourlas}.

The similarity between error-correcting codes of this type and Ising
spin systems was first pointed out by Sourlas\cite{Sourlas}, who
formulated the mapping of a simpler code, somewhat similar to the
one presented here, onto an Ising spin system Hamiltonian. We
recently extended the work of Sourlas, that focused on extensively connected
systems, to the finite connectivity case\cite{us_sourlas}.

To facilitate the current investigation we first map the problem to
that of an Ising model with finite connectivity. We employ the binary
representation $(\pm1)$ of the dynamical variables $\bS$ and $\btau$
and of the vectors $\bJ$ and $\JO$ rather than the Boolean $(0,1)$
one; the vector $\JO$ is generated by taking products of the relevant
binary message bits $J^{0}_{\left\langle i_{1}, i_{2} \ldots
\right\rangle} \! = \!  \xi_{i_{1}} \xi_{i_{2}} \ldots $, where the
indices $i_{1},i_{2}\ldots $ correspond to the non-zero elements of
$B^{-1}A$, producing a binary version of $\JO$. As we use statistical
mechanics techniques, we consider the message and codeword
dimensionality ($N$ and $M$ respectively) to be infinite, keeping the
ratio between them $R \!=\!  N/M$, which constitutes the code rate,
finite. Using the thermodynamic limit is quite natural as
Gallager-type codes are usually used for transmitting long
($10^{4}\!-\!10^{5}$) messages, where finite size corrections are
likely to be negligible.  To explore the system's capabilities we
examine the Hamiltonian
\begin{eqnarray}
\label{eq:Hamiltonian}
{\cal H} &=& \!\!\!\!\! \sum_{<i_1,..,i_K;j_1,..,j_L>}
      \mbox{\hspace*{-5mm}} \cD_{<i_1,..,i_K;j_1,..,j_L>} \ \delta
      \biggl[-1 \ ; \ \cJ_{<i_1,..,i_K;j_1,..,j_L>} \nonumber \\
      &\cdot & S_{i_1}\ldots S_{i_K} \tau_{j_1}\ldots\tau_{j_L}
      \biggr] - \frac{F_s}{\beta} \sum_{i=1}^{N} S_i -
      \frac{F_{\tau}}{\beta} \sum_{j=1}^{M} \tau_j \ .
\end{eqnarray}
The tensor product $\cD_{<i_1,..,i_K;j_1,..,j_L>}
\cJ_{<i_1,..,i_K;j_1,..,j_L>}$, where $\cJ_{<i_1,..,j_L>} \! = \!
\xi_{i_{1}} \xi_{i_{2}} \ldots \xi_{i_{K}} \zeta_{j_{1}} \zeta_{j_{2}}
\ldots \zeta_{j_{L}}$, is the binary equivalent of $A\bxi \!  + \!
B\bz$, treating both signal ($\bS$ and index $i$) and noise ($\btau$
and index $j$) simultaneously.  Elements of the sparse connectivity
tensor $\cD_{<i_1,..,j_L>}$ take the value 1 if the corresponding
indices of both signal and noise are chosen (i.e., if all
corresponding indices of the matrices $A$ and $B$ are 1) and 0
otherwise; it has $C$ unit elements per $i$-index and $L$ per
$j$-index representing the system's degree of connectivity.  The
$\delta$ function provides $1$ if the selected sites' product
$S_{i_1}\ldots S_{i_K} \tau_{j_1}\ldots\tau_{j_L}$ is in disagreement
with the corresponding element $\cJ_{<i_1,..,j_L>}$, recording an
error, and $0$ otherwise. Notice that this term is not frustrated, as
there are $M\! +\!N$ degrees of freedom and only $M$ constraints from
Eq.(\ref{eq:decoding}), and can therefore vanish at sufficiently low
temperatures. The last two terms on the right represent our prior
knowledge in the case of sparse or biased messages $F_s$ and of the
noise level $F_{\tau}$ and require assigning certain values to these
additive fields. The choice of $\beta\!  \rightarrow \!  \infty$
imposes the restriction of Eq.(\ref{eq:decoding}), limiting the
solutions to those for which the first term of
Eq.(\ref{eq:Hamiltonian}) vanishes, while the last two terms, scaled
with $\beta$, survive. Note that the noise dynamical variables $\btau$
are irrelevant to measuring the retrieval success $
m = \frac{1}{N} \ \left\langle \sum_{i=1}^{N} \ \xi_{i} \
\mbox{sign} \left\langle S_{i} \right\rangle_{\beta} \right\rangle_{\xi} \ .$
The latter monitors the normalized mean overlap between the Bayes-optimal
retrieved message, shown to correspond to the alignment of
$\left\langle S_{i} \right\rangle_{\beta}$ to the nearest binary
value\cite{Sourlas}, and the original message; the subscript $\beta$
denotes thermal averaging.

Since the first part of Eq.(\ref{eq:Hamiltonian}) is invariant under
the transformations $S_{i} \!\rightarrow\!  S_{i} \xi_{i}$,
$\tau_{j}\! \rightarrow\! \tau_{j} \zeta_{j}$ and $\cJ_{<i_1,..,j_L>}
\!\rightarrow\!  \cJ_{<i_1,..,j_L>} \xi_{i_{1}} ..  \xi_{i_{K}}
\zeta_{j_{1}} \zeta_{j_{2}}.. \zeta_{j_{L}} \! =\! 1$, it would be
useful to decouple the correlation between the vectors $\bS$, $\btau$
and $\bxi$, $\bz$. Rewriting Eq.(\ref{eq:Hamiltonian}) one obtains a
similar expression apart from the last terms on the right which become
$ F_s/\beta \sum_{k} S_{k} \ \xi_{k}$ and $ F_{\tau}/\beta \sum_{k}
\tau_{k} \ \zeta_{k}$.

The random selection of elements in $\cD$ introduces disorder to the
system which is treated via methods of statistical physics. More
specifically, we calculate the partition function ${\cal Z}
({\cD},\mbox{\boldmath $J$}) = \mbox{Tr}_{\{\bS,\btau\}} \exp [-\beta
{\cal H}]$ averaged over the disorder and the statistical properties
of the message and noise, using the replica
method\cite{us_sourlas,Wong_Sherrington,Dedom}.  Taking $\beta
\!\rightarrow\!\infty$ gives rise to a set of order parameters
\begin{eqnarray}
\label{eq:order_parameters}
q_{\alpha, \beta,.., \gamma} &=& \left\langle \frac{1}{N} \sum_{i=1}^{N}
Z_{i} \ S_{i}^{\alpha} \ S_{i}^{\beta},..,S_{i}^{\gamma}
\right\rangle_{\beta\rightarrow\infty} \nonumber \\ r_{\alpha, \beta,..,
\gamma} &=& \left\langle \frac{1}{M} \sum_{i=1}^{M} Y_{j} \
\tau_{j}^{\alpha} \ \tau_{j}^{\beta},.., \tau_{j}^{\gamma}
\right\rangle_{\beta\rightarrow\infty}
\end{eqnarray}
where $\alpha$, $\beta,..$ represent replica indices, and 
the variables $Z_{i}$ and $ Y_{j}$ come from enforcing the restriction 
of $C$ and $L$ connections per index respectively\cite{us_sourlas}:
\begin{equation}
\label{eq:delta}
\delta \left( \sum_{\left\langle i_{2},..,i_{K}
\right\rangle} \!\!\!\!\!\! \cD_{<i, i_{2},..,j_L>} -
C \right) = \oint_{0}^{2 \pi} \frac{d Z}{2 \pi} \ 
Z^{\sum_{\left\langle i_{2},.., i_{K} \right\rangle}
\!\!\! \cD_{<i, i_2,..,j_L>} -(C+1)} \ ,
\end{equation}
and similarly for the restriction on the $j$ indices.

To proceed with the calculation one has to make an assumption about
the order parameters symmetry. The assumption made here, and validated
later on, is that of replica symmetry in the following representation
of the order parameters and the related conjugate variables
\begin{eqnarray}
\label{eq:order_parameters_RS}
q_{\alpha, \beta.. \gamma} &=& a_{q} \int d x \ \pi(x) \
x^{l} \ , \ \widehat{q}_{\alpha, \beta.. \gamma}
= a_{\widehat{q}} \int d \hat{x} \ \widehat{\pi}(\hat{x}) \ 
\hat{x}^{l} \\ r_{\alpha, \beta.. \gamma} &=& a_{r}
\int d y \ \rho(y) \ y^{l}  \ , \ \widehat{r}_{\alpha,
\beta.. \gamma} = a_{\widehat{r}} \int d \hat{y} \
\widehat{\rho}(\hat{y}) \ \hat{y}^{l} \ , \nonumber
\end{eqnarray}
where $l$ is the number of replica indices, $a_{*}$ are normalization
coefficients, and $\pi(x), \widehat{\pi}(\hat{x}), \rho(y)$ and
$\widehat{\rho}(\hat{y})$ represent probability distributions.
Unspecified integrals are over the range $[-1,+1]$. One then obtains
an expression for the free energy per spin expressed in terms of these
probability distributions
\begin{eqnarray}
\label{eq:free_energy}
&\frac{1}{N}& \left\langle \ln {\cal Z} \right\rangle_{\xi,\zeta,\cD}
= \mbox{Extr}_{\{ \pi,\widehat{\pi},\rho,\widehat{\rho} \}} \biggl\{
 \frac{C}{K} \int \left[ \prod_{k=1}^{K} dx_{k} \
\pi(x_{k}) \right] \left[ \prod_{l=1}^{L} dy_{l} \ \rho(y_{l}) \right]
\ln \left[ 1 + \prod_{k=1}^{K} x_{k} \prod_{l=1}^{L} y_{l} \right]
\nonumber \\ &-& C \int dx \ d\hat{x} \ \pi(x) \
\widehat{\pi}(\hat{x}) \ \ln \left[ 1 + x \hat{x} \right] -
\frac{CL}{K} \int dy \ d\hat{y} \ \rho(y) \ \widehat{\rho}(\hat{y}) \
\ln \left[ 1 + y \hat{y} \right] \\ &+& \int \left[ \prod_{k=1}^{C}
dx_{k} \  \widehat{\pi}(\hat{x}_{k}) \right]
\left\langle \ln \left[\prod_{k=1}^{C} \left(1+\hat{x}_{k} \right) \
e^{F_{s}\xi} + \prod_{k=1}^{C} \left(1-\hat{x}_{k} \right) \
e^{-F_{s}\xi} \right] \right\rangle_{\xi} \nonumber \\ &+& \frac{C}{K}
\int \left[ \prod_{l=1}^{L}  d\hat{y}_{l} \
\widehat{\rho}(\hat{y}_{l}) \right] \left\langle \ln
\left[\prod_{l=1}^{L} \left(1+\hat{y}_{l} \right) \ e^{F_{\tau}\zeta}
+ \prod_{l=1}^{L} \left(1-\hat{y}_{l} \right) \ e^{-F_{\tau}\zeta}
\right] \right\rangle_{\zeta} -\frac{C}{K} \ln 2 \biggr\}\nonumber \ ,
\end{eqnarray}
where $\langle\cdot\rangle_{\xi}$ and $\langle\cdot\rangle_{\zeta}$ denote
averages over the input and noise distributions of the form
\begin{equation}
\langle\cdot\rangle_{\xi} = \sum_{\xi=\pm 1} \left\{ \frac{1+\tanh
F_{s}}{2} \ \delta_{\xi, -1} + \frac{1-\tanh F_{s}}{2} \ \delta_{\xi, 1}
\right\} (\cdot )
\end{equation}
and similarly for $\langle\cdot\rangle_{\zeta}$ where $F_{s}$ is
replaced by $F_{\tau}$.

The free energy can then be calculated via the saddle point
method. Solving the equations obtained by varying
Eq.(\ref{eq:free_energy}) w.r.t the probability distributions $\pi(x),
\widehat{\pi}(\hat{x}), \rho(y)$ and $\widehat{\rho}(\hat{y})$, is
generally difficult.
 The solutions obtained in the case of unbiased messages
(the most interesting case as most messages are compressed prior to
transmission) are for the ferromagnetic phase:
\begin{eqnarray}
\label{eq:sol_ferro}
 \pi(x) &=& \delta (x-1) \ , \ 
\widehat{\pi}(\hat{x}) = \delta (\hat{x}-1) \nonumber \\
 \rho(y) &=& \delta (y-1) \ , \
\widehat{\rho}(\hat{y}) = \delta (\hat{y}-1) \ ,
\end{eqnarray}
and for the paramagnetic phase (there is no spin-glass solution due to
lack of frustration):
\begin{eqnarray}
\label{eq:sol_para}
 \pi(x) &=& \delta (x) \ , \ \widehat{\pi}(\hat{x}) = \delta
(\hat{x}) \ , \ \widehat{\rho}(\hat{y}) = \delta (\hat{y}) \nonumber
\\ \rho(y) &=& \frac{1+\tanh F_{\tau}}{2} \ \delta (y-\tanh F_{\tau})
+ \frac{1-\tanh F_{\tau}}{2} \ \delta (y+\tanh F_{\tau}) \ .
\end{eqnarray}

It is easy to verify that these solutions obey the saddle point
equations. However, it is necessary to validate the stability of the
solutions and the replica symmetric ansatz itself.  To address these
questions we obtained solutions to the system described by the
Hamiltonian (\ref{eq:Hamiltonian}) via the TAP method of finitely
connected systems\cite{us_sourlas}; we solved the saddle point
equations derived from Eq.(\ref{eq:free_energy}) numerically,
representing all probability distributions by up to $10^4$ bin models
and by carrying out the integrations via Monte-Carlo methods; finally,
to show the consistency between theory and practice we carried out
large scale simulations for several cases, which will be presented
elsewhere.
The results obtained by the various methods are in complete
agreement.

The various methods indicate that the solutions may be divided to two
different categories characterized by $K\!=\! L\!=\! 2$ and by either
$K\!\ge\! 3$ or $L\!\ge\! 3$, which we therefore treat
separately.

For unbiased messages and either $K\!\ge\! 3$ or $L\!\ge\! 3$ we
obtain the solutions (\ref{eq:sol_ferro}) and (\ref{eq:sol_para}) both
by applying the TAP approach and by solving the saddle point equations
numerically. The former was carried out at the value of $F_{\tau}$
which corresponds to the true noise and input bias levels (for
unbiased messages $F_s\!=\! 0$) and thus to Nishimori's
condition\cite{Nishimori}, where no replica symmetry breaking effect
is expected.  This is equivalent to having the correct prior within
the Bayesian framework\cite{Sourlas_EPL} and enables one to obtain
analytic expressions for some observables as long as some gauge
requirements are obeyed\cite{Nishimori}. Numerical solutions show the
emergence of stable dominant delta peaks, consistent with those of
(\ref{eq:sol_ferro}) and (\ref{eq:sol_para}). The question of
longitudinal mode stability (corresponding to the replica symmetric
solution) was addressed by setting initial conditions for the
numerical solutions close to the solutions (\ref{eq:sol_ferro}) and
(\ref{eq:sol_para}), showing that they converge back to these
solutions which are therefore stable.

The most interesting quantity to examine is the maximal code rate, for
a given corruption process, for which messages can be perfectly
retrieved.  This is defined in the case of $K,L\!\ge\! 3$ by the value
of $R \! = \! K/C \! = \! N/M$ for which the free energy of the
ferromagnetic solution becomes smaller than that of the paramagnetic
solution, constituting a first order phase transition. A schematic
description of the solutions obtained is shown in the inset of
Fig.1a. The paramagnetic solution ($m\!=\!0$) has a lower free energy
than the ferromagnetic one (low/high free energies are denoted by the
thick and thin lines respectively, there are no axis lines at
$m\!=\!0,1$) for noise levels $p\! >\! p_c$ and vice versa for $p\!
\le \! p_c$; both solutions are stable. The critical code rate is
derived by equating the ferromagnetic and paramagnetic free energies
to obtain
\begin{equation}
R_{c}\!=\! 1\!-\!H_{2}(p)\!=\! 1\!+\!\left(p \log_{2} p \!+\!(1-p)
\log_{2} (1-p) \right) \ .
\end{equation}
This coincides with {\em Shannon's capacity}.  To validate these
results we obtained TAP solutions for the unbiased message case
($K\!=\! L\!=\! 3$, $C\!=\!6$). Averages over 10 solutions obtained
for different initial conditions in the vicinity of the stable
solutions are presented in Fig.1a (as $+$) in comparison to Shannon's
capacity (solid line).

Analytical solutions for the saddle point equations cannot be obtained
for the case of biased patterns and we therefore resort to numerical
methods and the TAP approach. The maximal information rate (i.e., code
rate $\times H_2(f_{s}=(1+\tanh F_{s})/2)$ - the source redundancy)
obtained by the TAP method ($\Diamond$) and numerical solutions of the
saddle point equations ($\Box$), averaged for each noise level over
solutions obtained for 10 different starting points in the vicinity of
the analytical solution, are shown in Fig.1a. Numerical results have
been obtained using $10^3 \!-\! 10^4$ bin models for each probability
distribution and had been run for $10^{5}$ steps per noise level
point. The various results are highly consistent and practically
saturate Shannon's bound for the same noise level.

The MN code for $K,L \ge 3$ seems to offer optimal performance.
However, the main drawback is rooted in the co-existence of the stable
$m=1$ and $m=0$ solutions, shown in Fig.1a (inset), which implies that
from some initial conditions the system will converge to the undesired
paramagnetic solution. Moreover, studying the ferromagnetic solution
numerically shows a highly limited basin of attraction, which becomes
smaller as $K$ and $L$ increase, while the paramagnetic solution at
$m=0$ {\em always} enjoys a wide basin of attraction. As initial
conditions for the decoding process are typically of close-to-zero
magnetization (almost no prior information about the original message
is assumed) it is highly likely that the decoding process will
converge to the paramagnetic solution. This performance has been
observed via computer simulations by us and by others\cite{MacKay}.

While all codes with $K,L \ge 3$ saturate Shannon's bound and are
characterized by a first order, paramagnetic to ferromagnetic, phase
transition, codes with $K\!=\! L\!=\!  2$ show lower performance and
different physical characteristics.  The analytical solutions
(\ref{eq:sol_ferro}) and (\ref{eq:sol_para}) are unstable at some flip
rate levels and one resorts to solving the saddle point equations
numerically and to TAP based solutions. The picture that emerges is
sketched in the inset of Fig.1b: The paramagnetic solution dominates
the high flip rate regime (appearing as a dominant delta peak in the
numerical solutions) up to the point $p_{1}$ (denoted as 1 in the
inset) in which a stable, ferromagnetic solution, of higher free
energy, appears (thin lines at $m\!=\!\pm 1$). At a lower flip rate
value $p_{2}$ the paramagnetic solution becomes unstable (dashed line)
and is replaced by two stable sub-optimal ferromagnetic (broken
symmetry) solutions which appear as a couple of peaks in the various
probability distributions; typically, these have a lower free energy
than the ferromagnetic solution until $p_{3}$, after which the
ferromagnetic solution becomes dominant (at some code rate values it
is dominant directly following the disappearance of the paramagnetic
solution). Still, only once the sub-optimal ferromagnetic solutions
disappear, at the spinodal point $p_{s}$, a unique ferromagnetic
solution emerges as a single delta peak in the numerical results (plus
a mirror solution).  The point in which the sub-optimal ferromagnetic
solutions disappear constitutes the maximal practical flip rate for
the current code rate and was defined numerically ($\Diamond$) and via
TAP solutions ($+$) as shown in Fig.1b.

Notice that initial conditions for both TAP and the numerical
solutions were chosen almost randomly, with a very slight bias of
${\cal O}(10^{-12})$, in the initial magnetization. The TAP dynamical
equations are identical to those used for practical BP
decoding\cite{us_sourlas}, and therefore provide equivalent results to
computer simulations with the same parameterization, supporting the
analytical results.  The excellent convergence results obtained point
out the existence of a unique pair of global solutions to which the
system converges (below $p_{s}$) {\em from practically all initial
conditions}.  This observation and the practical implications of using
the $K\!=\! L\!=\!  2$ code have not been obtained by information
theory methods (e.g.\cite{MacKay}); these prove the existence of very
good codes for $C,L\!\ge\!  3$, and examine decoding properties
only via numerical simulations.

In this Letter we examined the typical performance of Gallager-type
codes. We discovered that for a certain choice of parameters, either
$K\!\ge\!  3$ or $L\!\ge\!  3$, one obtains optimal performance,
saturating Shannon's bound. This comes at the expense of a decreasing
basin of attraction making the decoding process increasingly
impractical.  Another code, $K\!=\! L\!=\!  2$, shows close to optimal
performance with a very large basin of attraction, making it highly
attractive for practical purposes.  Studying the typical performance
of Gallager-type codes, which complements the methods used in the
information theory literature, is the first step towards understanding
their exceptional performance and in the search for a principled
method for designing optimal Gallager-type codes.  Important aspects
that are yet to be investigated include other noise types, irregular
constructions and the significance of finite size effects.

{\small
\vspace*{0.1in} {\bf \hspace*{-1em} Acknowledgement} Support by the
JSPS RFTF program (YK), The Royal Society and EPSRC grant GR/L19232
(DS) is acknowledged.}

\newpage

\begin{figure}
\vspace*{1.5cm}
\begin{center} \leavevmode  \epsfysize =7.0cm 
\epsfbox[170 150 450 360]{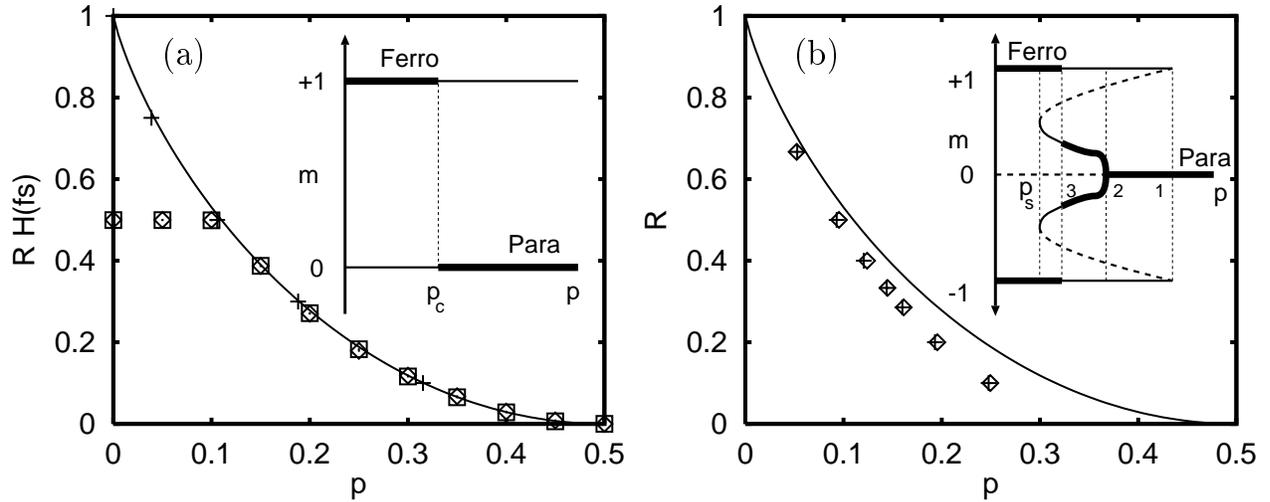}
\end{center}
\vspace*{2.5cm}
\caption{Critical code rate as a function of the flip rate $p$,
obtained from numerical solutions and the TAP approach ($N\!=\!
10^4$), and averaged over 10 different initial conditions with error
bars much smaller than the symbols size. (a) Numerical solutions for
$K\!  =\! L \!=\!3$, $C\!=\! 6$ and varying input bias $f_{s}$
($\Box$) and TAP solutions for both unbiased ($+$) and biased
($\Diamond$) messages; initial conditions were chosen close to the
analytical ones. The critical rate is multiplied by the source
information content to obtain the maximal information transmission
rate, which clearly does not go beyond $R\!=\!3/6$ in the case of
biased messages; for unbiased patterns $H_{2}(f_{s})\!=\!1$. Inset:
The ferromagnetic and paramagnetic solutions as functions of $p$;
thick and thin lines denote stable solutions of lower and higher free
energies respectively. (b) For the unbiased case of $K\!  =\! L
\!=\!2$; initial conditions for the TAP ($+$) and the numerical
solutions ($\Diamond$) are of almost zero magnetization. Inset: The
ferromagnetic (optimal/sub-optimal) and paramagnetic solutions as
functions of $p$; thick and thin lines are as in (a), dashed lines
correspond to unstable solutions.}
\end{figure}


\begin{thebibliography}{99}
\bibitem{Shannon} C.E.~Shannon, {\em Bell Sys.Tech.J.}, {\bf 27}, 379
(1948); {\bf 27}, 623 (1948).
\bibitem{Gallager} R.G.~Gallager, {\em IRE Trans.Info.Theory}, {\bf IT-8},
21 (1962).
\bibitem{MacKay} D.J.C.~MacKay, {\em IEEE Trans.IT}, {\bf 45}, 399
(1999).
\bibitem{Richardson} T.~Richardson, A.~Shokrollahi and R.~Urbanke 
unpublished (1999).
\bibitem{tap}  D.~Thouless, P.W.~Anderson and R.G.~Palmer, {\em Phil.~Mag.}, 
{\bf 35}, 593 (1977). 
\bibitem{us_sourlas} Y.~Kabashima and D.~Saad, {\em Euro.Phys.Lett.}, {\bf
44} 668 (1998) and {\bf 45} 97 (1999).
\bibitem{Sourlas} N.~Sourlas, {\em Nature}, {\bf 339}, 693 (1989).
\bibitem{Wong_Sherrington} K.Y.M.~Wong and D.~Sherrington, {\em J.Phys.A},
{\bf 20}, L793 (1987).
\bibitem{Dedom} C.~De~Dominicis and P.Mottishaw, {\em J.Phys.A},
{\bf 20}, L1267 (1987).
\bibitem{Nishimori} H.~Nishimori, {\em Prog.Theo.Phys.}, {\bf 66},
1169 (1981).
\bibitem{Sourlas_EPL} N.~Sourlas, {\em Euro.Phys.Lett.}, {\bf 25}, 159 (1994).
\end{thebibliography}
\end{document}